Viewpoint article

# High-entropy ceramic thin films; A case study on transition metal diborides

P. H. Mayrhofer, A. Kirnbauer, Ph. Ertelthaler, C. M. Koller

Materials Science and Technology, TU Wien, 1060-Vienna, Austria

**Abstract**

High-entropy materials often outperform their lower-entropy relatives in various aspects, such as thermal stability and fracture toughness. While there are extensive research activities in the field of high-entropy alloys, comparably little is performed for high-entropy ceramics, and especially for high-entropy diborides. Here we show, that not only the hardness of $ZrB_2$ layers can be improved from 43.2±1.0 GPa to 45.8±1.0 GPa to 47.2±1.8 GPa through the formation of solid solution ternary diborides ($Zr_{0.61}Ti_{0.39}B_2$) and high-entropy diborides ($Zr_{0.23}Ti_{0.20}Hf_{0.19}V_{0.14}Ta_{0.24}B_2$), respectively, but especially their thermal stability against structural rearrangements and decomposition towards the constituting binary diborides.

a)   electronic mail: paul.mayrhofer@tuwien.ac.at

Transition-metal (TM) borides are a special class of ultra-high temperature ceramics (UHTC). Among these TM compounds, refractory borides such as $ZrB_2$, $TiB_2$, $TaB_2$, and $HfB_2$ are attractive candidates for various applications such as cutting tools, molten metal containments, and microelectronics (e.g., as buffer layers) [1-4]. They are promising for many ultra-high temperature applications, because of their thermomechanical and chemical properties, their ultra-high melting temperatures up to ~4000 ºC, and excellent high temperature strengths. However, TM diborides have a comparably low fracture toughness and oxidation resistance, where especially the formation of volatile B–O compounds limits their stability in oxygen and/or moisture containing atmospheres. Of course, these properties can further be tailored by alloying with other TM borides as well as other refractory ceramics [5-14], or elements in general. For example, the combination with Mo and Si allows the formation of glassy-like borosilicate oxides exhibiting self-healing abilities as well as self-lubricating properties at specific temperature windows [15].

When it comes to alloying of materials, phrases such as multinary or even high-entropy alloys (HEA) are often used [16,17]. While there is no exact criterion for multinary alloys, high-entropy alloys are commonly classified after their composition or entropy [18]. According to the composition-based definition, high-entropy alloys need to be composed of five or more principal elements, in equimolar ratios or at least with contents between 5 and 35 at% [19]. This composition-based definition leads to a maximum configurational molar entropy S (simply estimated after Boltzmann [20,21] with $S = k_B \sum_1^n x_i \cdot \ln x_i$, with $k_B$ being Boltzmann's constant and $x_i$ being the molar fraction of the i-th element) of 1.61R for an equimolar five-element alloy (with R being the ideal gas constant and $R = k_B N_A$, where $N_A$ is Avogadro's number). The extended composition-based definition (contents between 5 and 35 at%) would result in a minimum configurational entropy of 1.36R (for the five-element alloy). This could even be exceeded by equimolar four-element alloys, yielding S=1.39R. Furthermore, as the composition-based definition does not restrict the alloys to be single phased, the configurational entropy could be even smaller.

Therefore, a definition based on entropy itself (consequently, within a single phase) seems to be more correct for alloys named "high-entropy". With respect to entropy, the alloys are often divided into low (S<0.69R), medium (0.69R≤S<1.61R), and high (S≥1.61R) entropy alloys [22]. The minimum criterion for medium entropy (S = 0.69R) and high-entropy (S =1.61R) is obtained for single-phased equimolar two-element and five-element alloys, respectively. The combination of both definitions – composition-based and entropy-based – resulted in a well-accepted compromise of a minimum entropy of S≥1.5R [23] for high-entropy alloys, which actually excludes any alloy with fewer than 5 elements. Accordingly, low entropy alloys and medium entropy alloys are often defined with S<R and R≤S<1.5R, respectively [24]. None of the traditional multicomponent alloys (like highly alloyed steels, Ni- or Co-based superalloys) have S≥1.5R, making the definition "high-entropy alloys" for S≥1.5R valuable.

Alloyed modifications of ceramic materials such as nitrides, oxides, carbides, and borides, can be defined as high-entropy ceramics (HEC) – combining the above-mentioned definition with the definition mentioned in [24] – if they are single-phased and composed of at least five corresponding binary nitrides, oxides, carbides, and borides, respectively, having S≥1.5R. Their configurational entropy (i.e., mixing entropy) now considers not elements, but the constituting binary compounds (in our case here, diborides). In this regard, we want to mention that the entropy now is defined per mole formula unit (e.g., TiN, $Al_2O_3$, TaC, $ZrB_2$). Contrary to HEAs where the entropy can directly (including Avogadro's number $N_A$) be given per atom as well, for HEC probably one entire sublattice (e.g., the nitrogen or the boron sublattice) is unchanged with respect to the constituting binaries. Therefore, presenting the mixing entropy of ceramics per atom, needs careful consideration of the formula unit (fu) and if all sublattices need to be considered or not [25]. It may be wise to use the above-mentioned presentation per (mole) formula unit. This needs to be considered when comparing the mixing entropy ($\Delta S_{mix}$, where the configurational entropy provides the strongest contribution [26]) with the mixing enthalpy ($\Delta H_{mix}$). This is especially relevant when estimating the mixing free energy ($\Delta G_{mix} = \Delta H_{mix} - T \cdot \Delta S_{mix}$), which clearly suggests that materials having large $\Delta S_{mix}$ (hence large configurational entropy) significantly lower the mixing free energy, especially with increasing temperature. Thus, also element combinations yielding positive mixing enthalpies, which have a tendency to phase separate, can be stabilized through the entropy term ($T \cdot \Delta S_{mix}$) at higher temperatures.

While there are extensive research activities with respect to HEAs, comparably little is performed for HECs, although for both, the same four core-effects (high-entropy; lattice distortion; sluggish diffusion; and cocktail effect [22]) provide a huge potential to increase their performance. With respect to HEC thin films, most activities concentrate on nitride and carbide coatings [27]. However, ground breaking results are missing (contrary to HEAs), most likely because fracture toughness and thermal stability are rarely investigated. Contrary to nitrides and carbides, there is almost no report about high-entropy boride (HEB) thin films; although, major application fields for borides are related with high temperatures, where the four core-effects of high-entropy materials (HEM) can be extremely beneficial. Recently, outstanding properties were reported for high-entropy metal diborides (in bulk form) [28]. As HEAs often significantly outperform their lower-entropy relatives with respect to fracture toughness [29], and especially transition metal diborides are infamous for their brittleness, they are excellent candidates for detailed studies on this high-entropy effect as well.

Here, we use non-reactive magnetron sputtering to prepare $ZrB_2$, $(Zr,Ti)B_2$, and $(Zr,V,Ti,Ta,Hf)B_2$ coatings, to investigate the potential of $TMB_2$ to form HECs, and to study the role of alloying on structure and mechanical properties, as well as thermal stabilities. The corresponding $TMB_2$ crystallize in the $AlB_2$ prototype structure, hexagonal C32 with space group P6/mmm, in which B is located in the interstices between (0001) close-packed planes of metal species. Further details of these $TMB_2$ are listed in Ref. [30]. All coatings prepared and studied exhibit hardnesses above 40 GPa in their as-deposited condition.

$ZrB_2$-based thin films, ~5 µm thick, are grown on austenitic stainless steel and sapphire ($1\bar{1}02$) substrates at 450 °C by magnetically-unbalanced magnetron sputter deposition from a stoichiometric $ZrB_2$ target (75 mm in diameter, 99.5 % purity, Plansee Composite Materials GmbH) in Ar (99.999 % purity) discharges, using a modified Leybold Heraeus Z400 system with a base pressure below 0.4 mPa ($4\times10^{-6}$ mbar). The substrates are parallel to the target and separated by 5 cm (facing the center of the target-race-track). The incident metal (Me) flux $J_{Me}$ is estimated based upon measurements of the deposition rate, film composition (see next paragraph), film thickness, and assuming bulk density. The ion flux $J_{Ar+}$ and the ion energy $E_{Ar+}$ bombarding the growing film are determined using Langmuir-probe measurements. The target power density was 4.4 W·cm$^{-2}$ (yielding a deposition rate of ~3.6 µm·h$^{-1}$) while using an Ar flow of 30 sccm, resulting in a pressure of 0.38 Pa and a $J_{Ar+}/J_{Me}$ ratio of 2.6. $E_{Ar+}$ is maintained constant at ~30 eV in these experiments by applying a substrate bias of -50 V. Experiments for $ZrB_2$ with 20 sccm, 30 sccm, and 40 sccm Ar flow, yielded $J_{Ar+}/J_{Me}$ ratios of 2.8, 2.6, and 2.4, respectively. As 20 sccm Ar flow resulted in unstable deposition conditions, we used 30 sccm Ar flow for the thin films prepared here, as this provides the highest $J_{Ar+}/J_{Me}$ ratio for stable deposition. The ternary $(Zr,Ti)B_2$ and multinary $(Zr,Ti,Hf,V,Ta)B_2$ coatings are prepared by placing stoichiometric $TiB_2$, $HfB_2$, $VB_2$, and $TaB_2$ platelets (about 1x1x0.5 mm$^3$, 99.5% purity, Plansee Composite Materials GmbH) onto the race track of the $ZrB_2$ target.

Film compositions are measured by energy dispersive X-ray spectroscopy (EDS) calibrated using a $TiB_2$ standard whose composition was determined by elastic recoil detection analyses.

Indentation hardness H and indentation modulus E of the films on sapphire substrates are obtained by computer controlled nanoindentation measurements (using a Ultra-Micro-Indentation System equipped with a Berkovich diamond tip) with maximum loads between 5 and 20 mN (in steps of 0.5 mN) for which the indentation depth is $\leq$ 10% of the film thickness. We calculated all H and E values according to the method proposed by Oliver and Pharr [31].

The microstructures of the layers in both the as-deposited and annealed states are determined using Bragg-Brentano x-ray diffraction (XRD) with Cu K$\alpha$ radiation and scanning electron microscopy (SEM). As-deposited films are annealed in vacuum (pressures $\leq 10^{-3}$ Pa) at temperatures $T_a$ of 1100 °C and 1500 °C for 10 min. To minimize any cross-diffusion effects (between substrate and layers) annealing experiments are only performed for layers on sapphire substrates.

B/Zr ratios in as-deposited $ZrB_2$ films are with 2.33, 2.37, and 2.34 for Ar flow rates of 20 sccm, 30 sccm, and 40 sccm within the error of the EDS measurements, respectively. The $(Zr,Ti)B_2$ and $(Zr,Ti,Hf,V,Ta)B_2$ coatings have slightly lower B/Me ratios of 2.29 and 2.08, respectively. Their metal atomic fractions are 60.5±0.5 % Zr and 39.5±0.2 % Ti respectively 22.9±0.3 % Zr, 20.2±0.2 % Ti, 19.0±0.3 % Hf, 14.0±0.7 % V, and 24.2±1.3 % Ta. Thus, we refer to our boride coatings as diborides with $ZrB_2$, $Zr_{0.61}Ti_{0.39}B_2$ [simply $(Zr,Ti)B_2$], and $Zr_{0.23}Ti_{0.20}Hf_{0.19}V_{0.14}Ta_{0.24}B_2$ [simply $(Zr,Ti,Hf,V,Ta)B_2$]. Their molar mixing entropies are 0.67R (0.0578 meV/fu·K) and 1.59R (0.1373 meV/fu·K), which represents the $(Zr,Ti,Hf,V,Ta)B_2$ clearly as a high-entropy ceramic if single-phased crystalline. The mixing enthalpies, estimated from ab initio calculated energy of formation differences between the solid solutions and the constituting binary diborides, are ~0.035 eV/at (0.012 eV/fu) for $Zr_{0.61}Ti_{0.39}B_2$ and ~0.032 eV/at (0.011 eV/fu) for $Zr_{0.2}Ti_{0.2}Hf_{0.2}V_{0.2}Ta_{0.2}B_2$. This clearly suggests that already for temperatures above ~210 and 80 K the mixing entropy term (T·$\Delta S_{mix}$) overrules the mixing enthalpy of (Zr,Ti)B$_2$ and (Zr,Ti,Hf,V,Ta)B$_2$, respectively. Thus, these alloys should be stable against decomposition towards their constituting binary diborides.

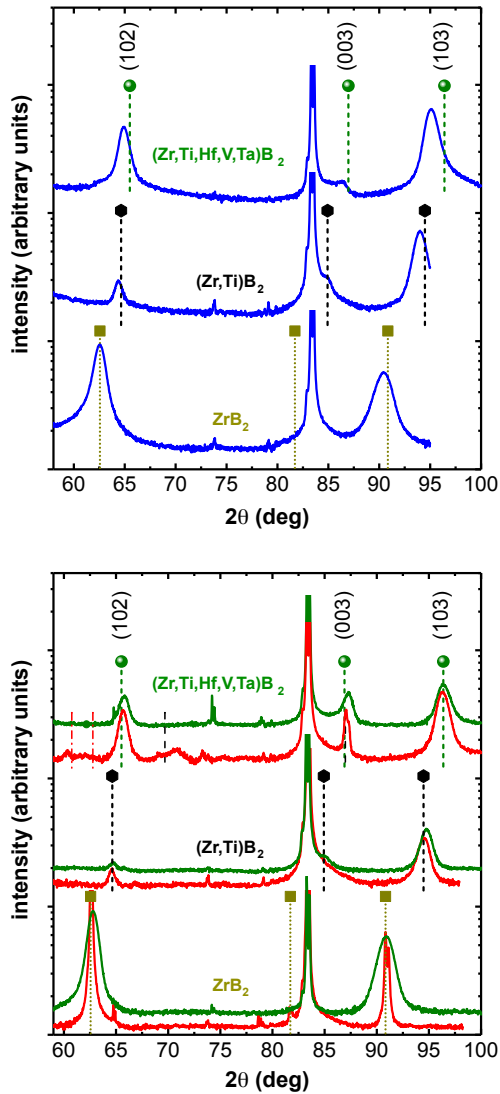

FIG. 1. XRD patterns from (a) as-deposited ZrB$_2$, Zr$_{0.61}$Ti$_{0.39}$B$_2$ [(Zr,Ti)B$_2$], and Zr$_{0.23}$Ti$_{0.20}$Hf$_{0.19}$V$_{0.14}$Ta$_{0.24}$B$_2$ [(Zr,Ti,Hf,V,Ta)B$_2$] thin films (blue patterns) and samples that have been vacuum-annealed (b) for 10 min at T$_a$ = 1100 °C (green patterns) and 1500 °C (red patterns). The (102), (003), and (103) XRD peak positions – calculated for a = 3.156 Å and c = 3.532 Å (ZrB$_2$), a = 3.095 Å and c = 3.420 Å [(Zr,Ti)B$_2$], and a = 3.095 Å and c = 3.360 Å [(Zr,Ti,Hf,V,Ta)B$_2$] – are marked.

The XRD peaks obtained from as-deposited ZrB$_2$, (Zr,Ti)B$_2$, and (Zr,Ti,Hf,V,Ta)B$_2$ layers over the 2θ range 20–105° suggest for the formation of single-phased TMB$_2$ solid solutions (see Fig. 1a, showing the coatings grown on sapphire substrates). Also on polycrystalline steel substrates, their preferred growth orientations are (102) and (103). The corresponding XRD peak positions suggest lattice constants of a = 3.11 Å and c = 3.56 Å for ZrB$_2$, which decrease to a = 3.09 Å and c = 3.44 Å for (Zr,Ti)B$_2$, and to a = 3.09 Å and c = 3.40 Å for (Zr,Ti,Hf,V,Ta)B$_2$. An extremely simplified estimation of the lattice constants with a linear interpolation [32] of the respective TMB$_2$ constituents, provides a = 3.12 Å and c = 3.41 Å for Zr$_{0.61}$Ti$_{0.39}$B$_2$, and a = 3.09 Å and c = 3.33 Å for Zr$_{0.23}$Ti$_{0.20}$Hf$_{0.19}$V$_{0.14}$Ta$_{0.24}$B$_2$. The surprisingly excellent agreement proofs the single phase solid solution nature of our coatings, and in combination with S=1.59R (previous paragraph) this defines our (Zr,Ti,Hf,V,Ta)B$_2$ as a high-entropy ceramic or more specific as a high-entropy diboride (HEB$_2$). Furthermore, also based on the lattice parameter differences (maximum between ZrB$_2$ and VB$_2$ of 5% for a and 14% for c), solid solutions are expected [33,34]. All binaries exhibit lattice parameter differences ≤8% (≤3% for a and ≤8% for c) from the linear interpolated lattice parameters of the (Zr,Ti,Hf,V,Ta)B$_2$ solid solution.

The three diborides exhibit a similar, very dense and nearly featureless growth morphology, see Figs. 2a, b, and c showing SEM fracture cross sections of ZrB$_2$, Zr$_{0.61}$Ti$_{0.39}$B$_2$, and Zr$_{0.23}$Ti$_{0.20}$Hf$_{0.19}$V$_{0.14}$Ta$_{0.24}$B$_2$, respectively. But the ternary Zr$_{0.61}$Ti$_{0.39}$B$_2$ shows a slightly lower deposition rate, which we account to the lower sputtering yield of TiB$_2$.

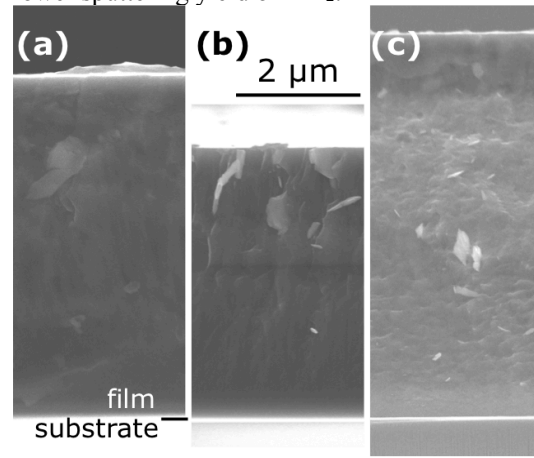

FIG. 2. Fracture cross sectional SEM images of as-deposited (a) ZrB$_2$, (b) Zr$_{0.61}$Ti$_{0.39}$B$_2$ [(Zr,Ti)B$_2$], and (c) Zr$_{0.23}$Ti$_{0.20}$Hf$_{0.19}$V$_{0.14}$Ta$_{0.24}$B$_2$ (HEB$_2$).

Even after annealing at 1500 °C (in vacuum), no separation of the solid solutions towards their binary diborides can be detected, see the red XRD patterns in Fig. 1b. The corresponding solid solution TMB$_2$ XRD peaks are still dominating, and also their preferred growth orientation is not changed, suggesting that no significant recrystallization effects occurred. But especially for (Zr,Ti)B$_2$ and (Zr,Ti,Hf,V,Ta)B$_2$, multiple small XRD peaks, in addition to the solid solution related ones, can be detected. When annealing the coating at 1100 °C, no such additional XRD peaks are detectable, see the green XRD patterns of (Zr,Ti)B$_2$ and (Zr,Ti,Hf,V,Ta)B$_2$. We address these additional small XRD peaks to the formation of lower boron containing relatives (mainly orthorhombic TM$_3$B$_4$ and TMB), triggered by B-loss at temperatures above 1100 °C (next paragraph). This is in agreement with our previous studies on Ti-B-N coatings, yielding especially for $T_a \geq 1000$ °C significant B-loss (due to the formation of volatile $H_3BO_3$ in the presence of residual oxygen and moisture) [35].

Chemical investigations of our coatings after (vacuum) annealing at 1100 °C and 1500 °C, also show significant B-loss with increasing $T_a$, see the decreasing B/Me ratios in Fig. 3a. Whereas $ZrB_2$ and $(Zr,Ti)B_2$ still exhibit B/Me ratios close to 2, even after annealing at $T_a = 1500$ °C, the B/Me ratio of our high-entropy diboride decreased significantly for $T_a > 1100$ °C to ~1.5 (from the as-deposited value of 2.09). Therefore, the additional small XRD peaks (present in the patterns after annealing at 1500 °C) are most pronounced for the high-entropy diboride, $(Zr,Ti,Hf,V,Ta)B_2$. Simultaneously, also the O content of our coatings increases (which was below the detection limit in the as-deposited state) with increasing $T_a$, where after annealing at 1500 °C the high-entropy diboride yields the highest O-content, see Fig. 3b. We envision, that especially in high-entropy ceramics (such as our $HEB_2$), where actually one sublattice can still be unchanged, the sluggish diffusion effect (one of the core-effects of HEAs) basically concentrates on the high-entropy sublattice. But the non-metal sublattice could even show increased diffusivities (due to the distortion of the metal sublattice), as suggested by our investigations.

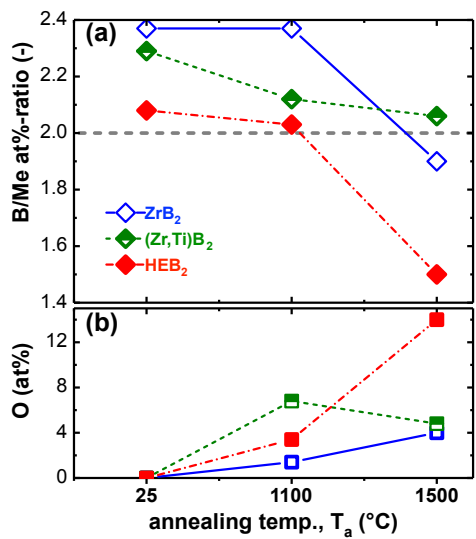

FIG. 3. (a) Boron to metal ratio (B/Me) of as-deposited $ZrB_2$, $Zr_{0.61}Ti_{0.39}B_2$ [$(Zr,Ti)B_2$], and $Zr_{0.23}Ti_{0.20}Hf_{0.19}V_{0.14}Ta_{0.24}B_2$ ($HEB_2$) and samples that have been vacuum-annealed for 10 min at $T_a = 1100$ and 1500 °C. Their O content is given in (b).

In addition to these chemical changes, the increased diffusivities during thermal annealing commonly lead to atomic rearrangements of structural built-in defects (generated by the deposition process) to lower energy sites (including defect annihilation) resulting in relaxed lattices (i.e., recovered structures) [36,37]. Thus, diffraction peak widths Γ – an effective measure for the presence and amount of microstructural defects [38,39] – of stoichiometric binary nitride and carbide layers decrease when (vacuum) annealed at temperatures above their deposition temperature [40-43]. Typically, already binary diborides exhibit a higher thermal stability during vacuum annealing, due to their stronger bonds. This is even more pronounced if additional nanostructural features are present like the excess-boron-stabilized nanoclumns of $TiB_{2.4}$ [44]. There, bundles of ~5-nm-diameter $TiB_2$ subcolumns are separated and encapsulated by an ultra-thin B-rich tissue phase. The thereby generated nanocolumnar structure is thermally stable to post-annealing temperatures of at least 700 °C (the maximum temperature investigated in [44]). As such structural defects are effective obstacles for dislocation glide and motion, not only the diffraction peak width Γ of $TiB_{2.4}$ remains constant with $T_a$, but also the hardness, which is determined by resistance to bond distortion and dislocation formation and motion during loading (e.g., indentation as in our case).

Interestingly, the full width at half maximum intensities Γ of the 102 and 103 reflections from as-deposited $ZrB_2$ are significantly larger than those obtained from as-deposited $(Zr,Ti)B_2$ and $(Zr,Ti,Hf,V,Ta)B_2$ layers, see Fig. 4a showing $\Gamma_{103}$. This would suggest for a smaller nanostructure or larger lattice microstrains [39] of $ZrB_2$, which is somehow counterintuitive when comparing this binary with the ternary and the high-entropy diboride. But, we basically account this to the higher B content of $ZrB_2$ ($ZrB_{2.37}$ when prepared with 30 sccm Ar flow, whereas $(Zr,Ti)B_2$ and $(Zr,Ti,Hf,V,Ta)B_2$ have B/Me ratios of 2.29 and 2.08), where excess B can promote a nanocolumnar structure comparable to the above-mentioned $TiB_{2.4}$ [44]. Generally, strengthening through nanostructures or lattice microstrains is less thermally stable than through solid solution mechanisms [33] present in our ternary $(Zr,Ti)B_2$ and high-entropy $(Zr,Ti,Hf,V,Ta)B_2$.

The full width at half maximum intensity $\Gamma_{103}$ of the 103 reflection significantly decreases from 1.60351 ° to 0.21837 ° for $ZrB_2$, but only marginally from 1.17062 ° to 1.13688 ° for $(Zr,Ti)B_2$, and from 1.24208 ° to 1.18195 ° for $(Zr,Ti,Hf,V,Ta)B_2$ during annealing at 1500 °C (Fig. 3a). The corresponding $\Gamma_{102}$ shows a similar behavior and decreases from 1.0257 ° to 0.31378 ° for $ZrB_2$, and only slightly from 0.70665 ° to 0.58761 ° for $(Zr,Ti)B_2$, and from 0.99614 ° to 0.86001 ° for $(Zr,Ti,Hf,V,Ta)B_2$. Thus, there are significant changes in either nanostructure or local lattice microstrains [39] for $ZrB_2$ during annealing at 1500 °C, but not for $(Zr,Ti)B_2$ and $(Zr,Ti,Hf,V,Ta)B_2$. Simultaneously, also the lattice constants of $ZrB_2$ approach their bulk values (increase in a to 3.156 Å and decrease in c to 3.532 Å), whereas those of $(Zr,Ti)B_2$ only slightly change to a = 3.095 Å and c = 3.420 Å, and those of $(Zr,Ti,Hf,V,Ta)B_2$ change to a = 3.095 Å and c = 3.360 Å, see Figs. 4b and c, respectively. Hence, also the macrostresses of $ZrB_2$ significantly decrease during annealing at 1500 °C, but those of (Zr,Ti)B$_2$ and (Zr,Ti,Hf,V,Ta)B$_2$ only slightly change. The ab initio obtained lattice parameters for Zr$_{0.2}$Ti$_{0.2}$Hf$_{0.2}$V$_{0.2}$Ta$_{0.2}$B$_2$ are with a = 3.095 Å, b = 3.097 Å, and c = 3.348 Å in excellent agreement with those of our high-entropy (Zr,Ti,Hf,V,Ta)B$_2$, especially after annealing at 1100 °C or 1500 °C.

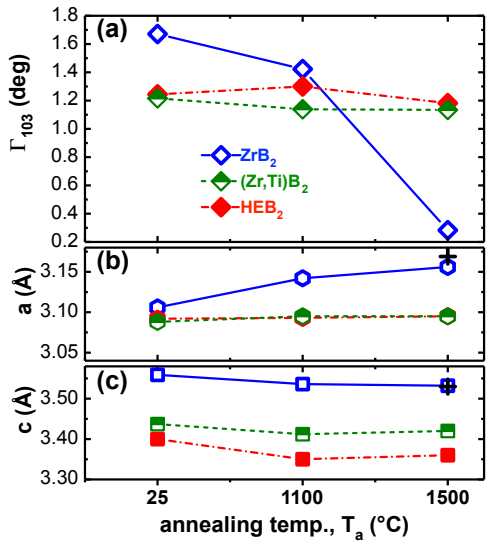

FIG. 4. (a) Full width at half maximum $\Gamma_{103}$ of the (103) XRD reflection from as-deposited ZrB$_2$, Zr$_{0.61}$Ti$_{0.39}$B$_2$ [(Zr,Ti)B$_2$], and Zr$_{0.23}$Ti$_{0.20}$Hf$_{0.19}$V$_{0.14}$Ta$_{0.24}$B$_2$ (HEB$_2$) and samples that have been vacuum-annealed for 10 min at T$_a$ = 1100 and 1500 °C. The corresponding lattice constants a and c are given in (b) and (c). For comparison, the lattice constants of bulk ZrB$_2$ are indicated by crosses (a = 3.169 Å and c = 3.530 Å, JCPDF 00-034-0423).

The hardness H and indentation modulus E of our as-deposited ZrB$_2$ layers (on sapphire) is 43.2±1.0 GPa and 543.1±15.2 GPa, respectively. On austenite substrates (same deposition run as used for the sapphire substrates), the ZrB$_2$ layers yield a similar hardness but lower indention modulus (H = 44.8±2.3 GPa and E = 466.6±15.8 GPa), due to the nanoindentation experiment although the indentation depth was below 10% of the coating thickness [45]. As-deposited, the (Zr,Ti)B$_2$ and (Zr,Ti,Hf,V,Ta)B$_2$ layers have higher hardnesses, H = 45.8±1.0 GPa and 47.2±1.8 GPa (Fig. 5a), but comparable E moduli, E = 537.6±12.2 GPa and 540.1±17.1 GPa (Fig. 5b), respectively. Consequently, the H/E ratio (Fig. 5c) is slightly higher for the ternary and high-entropy diborides.

After annealing at 1100 °C, the three coatings [ZrB$_2$, (Zr,Ti)B$_2$, and (Zr,Ti,Hf,V,Ta)B$_2$] exhibit comparable hardnesses (H ~42 GPa) and E-moduli (E ~545 GPa). The still high hardness values agree with the small structural changes (see Figs. 1 and 4) upon annealing at T$_a$ = 1100 °C. However, annealing at 1500 °C leads to a significant reduction in H for ZrB$_2$ (H ~28 GPa) but not for (Zr,Ti)B$_2$, which still exhibits ~36 GPa hardness. This agrees with the significant microstructural changes of ZrB$_2$ but the only minor changes of (Zr,Ti)B$_2$, represented by their XRD patterns and $\Gamma_{103}$, see Figs. 1 and 4. Especially after annealing at 1500 °C, the H/E ratio is significantly larger for (Zr,Ti)B$_2$ (H/E = 0.081) than for ZrB$_2$ (H/E = 0.057). Thus, we expect higher fracture toughness for (Zr,Ti)B$_2$.

The mechanical properties of the high-entropy diboride, after annealing at 1500 °C, could not be obtained due to the significant spallation from the sapphire substrate and the pronounced surface reaction with residual oxygen and moisture. The latter are represented by the significantly reduced B/Me ratio and rather high O-content (Fig. 3), leading to the formation of lower B-containing phases (Fig. 1b). The comparably still large $\Gamma_{103}$ values of their diboride phase suggests a similar thermal stability against decomposition and structural changes as obtained for (Zr,Ti)B$_2$. Consequently, the reaction with residual oxygen and moisture limits the huge potential of this HEB$_2$.

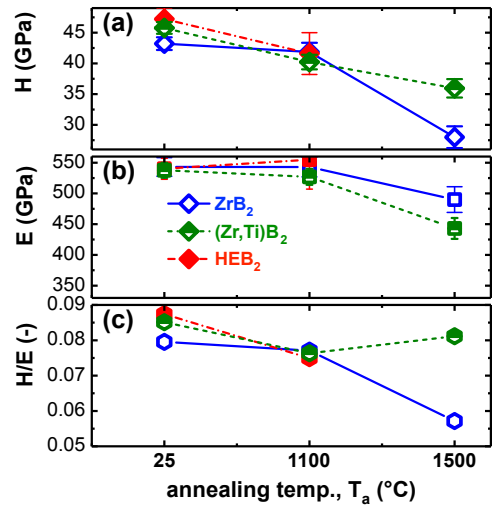

FIG. 5. (a) Hardness H and (b) indentation modulus E of as-deposited ZrB$_2$, Zr$_{0.61}$Ti$_{0.39}$B$_2$ [(Zr,Ti)B$_2$], and Zr$_{0.23}$Ti$_{0.20}$Hf$_{0.19}$V$_{0.14}$Ta$_{0.24}$B$_2$ (HEB$_2$) and samples that have been vacuum-annealed for 10 min at T$_a$ = 1100 and 1500 °C. The corresponding H/E ratios are given in (c).

Our results clearly show that high-entropy diborides, here (Zr,Ti,Hf,V,Ta)B$_2$, can successfully be prepared by non-reactive physical vapor deposition using the corresponding diboride composite targets. As-deposited, this HEB$_2$ (H = 47.2±1.8 GPa) is harder than the ternary (Zr,Ti)B$_2$ (H = 45.8±1.0 GPa) and binary ZrB$_2$ (H = 43.2±1.0 GPa). The high-entropy as well as the ternary diboride solid solutions are thermally stable against decomposition towards their constituting binary diborides, up to (vacuum) annealing temperatures of 1500 °C. But especially the HEB$_2$ shows a significant B-loss and O-uptake for T$_a$ > 1100 °C, limiting their huge potential and thermal stability. However, based on our results, we conclude that if B-loss and O-uptake can be limited (or even avoided), HEB$_2$ can guarantee high thermal stability against decomposition towards their binary or ternary diborides. Thus, also their hardness should be at least 36 GPa, even after annealing at 1500 °C, as obtained for the ternary (Zr,Ti)B$_2$.

Consequently, especially in thin film form (due to the limited volume available), developments of high-entropy alloys and high-entropy ceramics should consider also chemical variations by interactions with ambient atmosphere and materials. Furthermore, in high-entropy ceramics one sublattice can still be unchanged, and hence the four core-effects of HEAs (such as sluggish diffusion) basically concentrate on the high-entropy sublattice. But the non-metal sublattice could even show increased diffusivities (due to the distortion of the metal sublattice), as suggested by our investigations.

**Acknowledgements**

The authors acknowledge the use of the X-Ray Center and USTEM, TU Wien. The computational results presented have been achieved using the Vienna Scientific Cluster (VSC). The authors are also very grateful to Dr. D. Holec (Montanuniversität Leoben) for assistance with ab initio calculations and valuable discussions.

**Graphical Abstract**

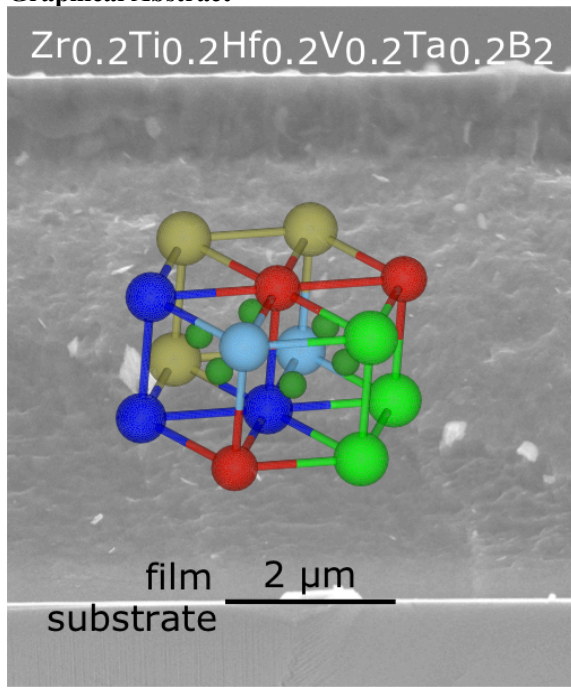
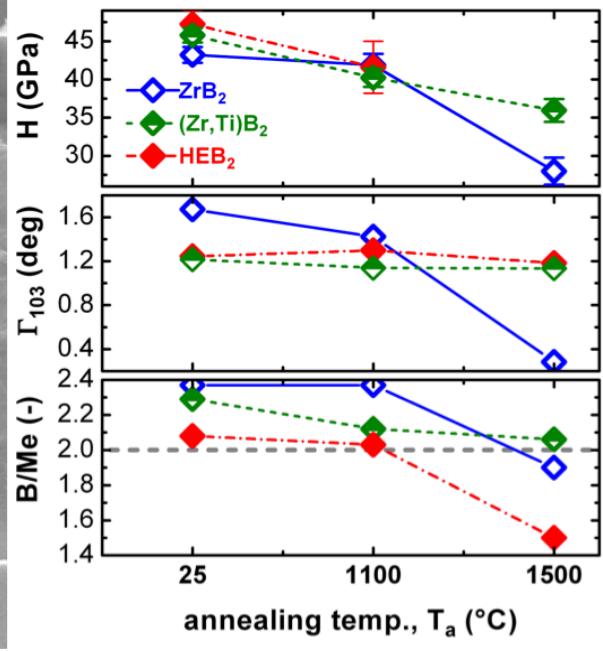